\documentclass[traditabstract,referee]{aa} 

\usepackage{natbib,graphicx,hyperref} 
\usepackage[varg]{txfonts}

\usepackage{amssymb}	

\begin{document}

\title{Solar activity: intrinsic periodicities \\beyond 11 years?}
\author{
  R.~H.~Cameron,\inst{1}
  \and
  M.~Sch{\"u}ssler\inst{1}
 }

\institute{Max-Planck-Institut f{\"u}r Sonnensystemforschung
Justus-von-Liebig-Weg 3, 37077 G{\"o}ttingen\\
\email{cameron@mps.mpg.de}
}

\keywords{Sun: activity -- Sun: magnetic fields -- Dynamo}

\titlerunning{Solar activity}

\date{Received ; Accepted}

\abstract {Power spectra of solar activity based on historical records
  of sunspot numbers and on cosmogenic isotopes show peaks with
  enhanced power apart from the dominant 11-year solar cycle, such as
  the 90-year Gleissberg cycle or the 210-year de Vries cycle. In a
  previous paper we have shown that the overall shape of the power
  spectrum is well represented by the results of the generic normal
  form model for a noisy and weakly nonlinear limit cycle, with
  parameters all determined by observations. Using this model as a
  null case, we show here that all local peaks with enhanced power,
  apart from the 11-year band, are consistent with realization
  noise. Even a $3\sigma$ peak is expected to occur with a probability
  of about 0.25 at least once among the 216 period bins resolved by the
  cosmogenic isotope data. This casts doubt upon interpretations of
  such peaks in terms of intrinsic periodicities of the solar dynamo
  process.}

\maketitle

\section{Introduction}
\label{sec:introduction}

Solar activity has been directly observed in terms of sunspot numbers
since about 400 years, while its level at earlier times can be inferred
from the concentrations of the cosmogenic isotopes $^{14}$C in tree
rings and $^{10}$Be in polar ice over the last $\sim$10,000 years. In
addition to the 11-year Schwabe cycle, these records show modulation of
the activity level on longer time scales. Power spectra show signal at
all periods, together with a number of peaks of locally enhanced power.
Peaks around 90 years (generally called the Gleissberg cycle), 210 years
(de Vries cycle), and perhaps 2400 years (Hallstatt cycle) of presumably
solar origin \citep{McCracken:etal:2013} have been discussed in the
literature \citep[see][and references therein]{Usoskin:2017}. Although
it is often assumed that these periodicities are intrinsic to the solar
dynamo process \citep[e.g.,][]{Beer:etal:2018}, in the absence of a
proper null case it has so far not been tested whether these power peaks
are consistent with noise due to stochastic random processes affecting the
dynamo. Even if such peaks are formally significant (say, at the
$3\sigma$ level), this does not necessarily exclude that they originate
from random noise. Here we propose an appropriate null case and test.

In \citet{2017ApJ...843..111C} we showed that the solar dynamo can be
reasonably described by a weakly nonlinear limit cycle affected by
random noise. Such a model is favored by stellar observations suggesting
that the Sun is near the critical rotation rate for which dynamo action
sets in \citep{vanSaders:etal:2016, Metcalfe:etal:2016,
  Metcalfe:etal:2017} and by solar observations of the scatter of active
region tilt angles \citep{2014ApJ...791....5J}, which are a basic
ingredient of Babcock-Leighton-type models of the solar dynamo
\citep{Charbonneau:2010, Cameron:Schuessler:2017}. This approach leads
to a normal-form model for a noisy system near a Hopf bifurcation, which
is generic in the sense that it does not depend on the detailed
properties of the system, such as its specific kind of nonlinearity. The
parameters of this model are all well constrained by solar
observations. Random time series resulting from the model show power
spectra consistent with the overall shape of the power spectrum of solar
activity derived from the sunspot record and cosmogenic isotopes,
including the occurrence and statistics of grand minima, i.e., extended
periods of very low activity \citep{2017ApJ...843..111C}.

These results suggest that the noisy normal-form model (henceforth
denoted as the NNF model) provides an adequate null case for evaluating
the significance of the various discrete periodicities (e.g., the
Gleissberg, de Vries, and Hallstatt cycles) that have been detected in
the power spectra based on cosmogenic isotope records,. In this paper,
we consider whether such peaks are consistent with realization noise
of the generic normal-form model.

\section{Data and Model}

We used the reconstructed record of solar activity based on cosmogenic
isotopes provided by \citet{Usoskin:etal:2016}.  To exclude a dependence
on the specific reconstruction we also employed the data of
\citet{Steinhilber:etal:2012} and found similar results. We compare
these empirical data with realizations of the NNF model
\citep{2017ApJ...843..111C} given, in terms of a stochastic differential
equation, by
\begin{eqnarray}
  \mathrm{d} X = \left(\beta+i\omega_0 -(\gamma_r + i \gamma_i) |X|^2
  \right)X \mathrm{d}t +\sigma X \mathrm{d}W_c \,,
\label{eqn:normalform}
\end{eqnarray}
where $X$ is a complex variable whose real and imaginary parts
represent, respectively, the poloidal and toroidal magnetic field
components.  As a measure of the activity level, we take
$|{\mathrm{Re}}(X)|$.  $\omega_0={2\pi}/{22}$~yrs$^{-1}$ is the linear
frequency of the magnetic cycle and the linear growth rate,
$\beta=1/50$~yrs$^{-1}$, is based on the recovery time from the
Maunder minimum.  The parameters ${\gamma }_{r}$ and ${\gamma }_{i}$
determine the cycle amplitude, $|X| =\sqrt{\beta /{\gamma }_{r}}$, and
the nonlinear cycle frequency, $\omega ={\omega }_{0}-{\gamma
}_{i}\beta /{\gamma }_{r}$. We assume the cycle frequency to be
unaffected by the (weak) nonlinearity, i.e. $\gamma_{i}=0$, and scale
the cycle amplitude in terms of the average group sunspot number since
1700, yielding ${\gamma}_{r}=\beta/64^2$.  For the noise term, $W_c$,
we consider a complex Wiener process with Gaussian distributed
increments and an amplitude of $\sigma=0.4/\sqrt{11}$~yrs$^{-1/2}$,
estimated from the variability of the observed cycle amplitudes.  For
our analysis, we computed a a set of 1000 different realizations of
the NNF model, each covering 10,000 years. For the numerical solution,
we used the Euler-Maruyama method with a timestep of 1 day.  Further
details can be found in \citet{2017ApJ...843..111C}.

\section{Results}

\subsection{Significance of single peaks}

\begin{figure}
	\includegraphics[width=\columnwidth]{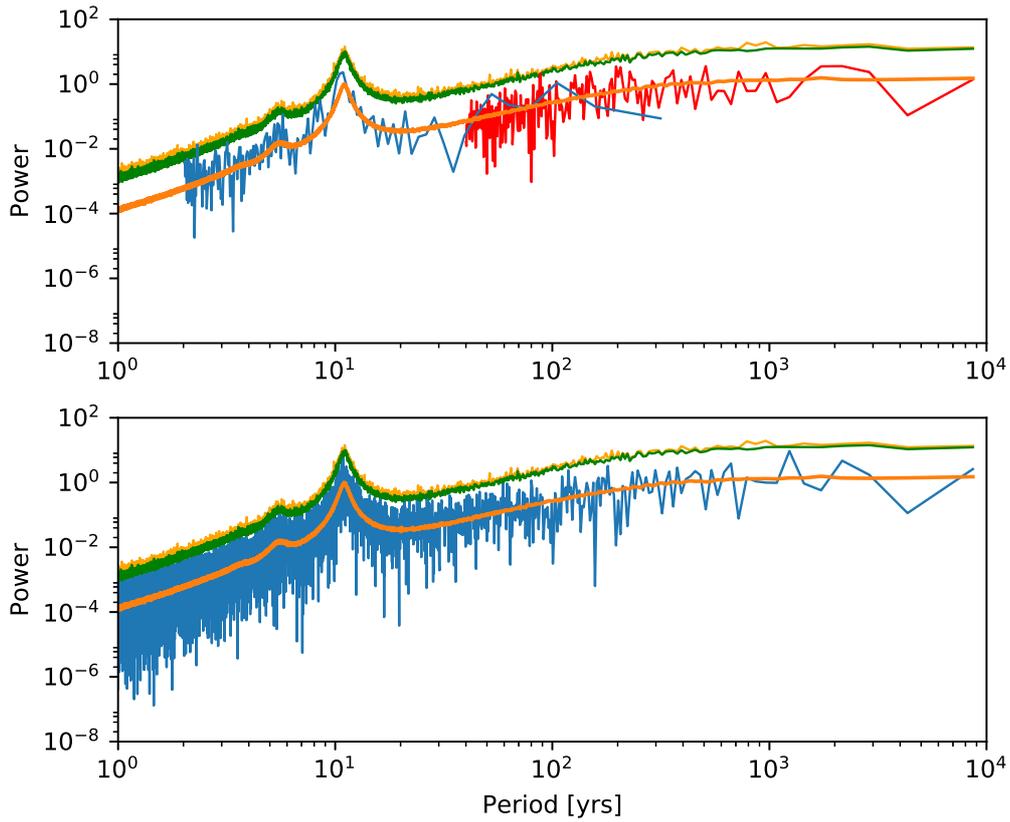}
        \caption{Observed and modelled power spectra of solar activity.
          {\it Upper panel:} Spectra based on the historical
          record of sunspot numbers (blue curve) and on the
          reconstruction from cosmogenic isotope data \citep[][red
            curve]{Usoskin:etal:2016}.  {\it Lower panel:} Spectrum
          obtained from one realization of the noisy normal form model
          \citep[NNF model][blue curve]{2017ApJ...843..111C}.  In both panels,
          the orange curve gives the median from 1000 realizations of
          the model, the green curve shows the corresponding 99.865\%
          ($3\sigma$) upper percentile, and the yellow curve represents
          the maximum of the 1000 realizations.}
    \label{fig:PS}
\end{figure}

 Estimating the significance of peaks in a power spectrum requires
  a null case.  Since the overall shape of the empirical spectrum
  based on cosmogenic isotopes is well represented by the NNF model
  \citep{2017ApJ...843..111C}, we propose this model as a
  physics-based null case instead of an ad-hoc choice such as red or
  white noise.  For the NNF model we can determine the standard
  deviation from the median spectrum (as well as the significance of
  local peaks) by considering a sufficient number of random
  realizations.
Figure~\ref{fig:PS} shows power spectra from the record of solar
activity (observed sunspot numbers%
\footnote{Version 2 of the yearly sunspot number:
http://www.sidc.be/silso/DATA/SN\_y\_tot\_V2.0.txt}
and inferred from cosmogenic isotope data) in the upper panel and from
one realization of the NNF model in the lower panel. In both panels,
the orange curve gives the median of the 1000 model realizations, the
green curve shows the (upper) 99.865\% percentile level from the set
of model realizations (corresponding to the $3\sigma$ level), and the
yellow curve displays the maximum power for each frequency point of
the 1000 realizations. 
The near coincidence of the
green and yellow curves indicates that, as expected on statistical
grounds, for 1000 realizations of a random process at least one peak
above the $3\sigma$ level in one of these realizations is expected for
any given frequency.  Considering the 216 periods over 40 years that
are resolved by the cosmogenic isotope data, the probability, $p$, of
finding at least one peak above the $3\sigma$ level in one realization
of the random model (with the same period resolution) is given by
\begin{equation}
p = 1 - 0.99865^{216} = 0.253\;,
\end{equation}
so that we expect such a peak in about a quarter of the realiations.
Therefore, the appearance of an outlier $3\sigma$ peak such as that
corresponding to the 88-year Gleissberg period in the
\citet{Usoskin:etal:2016} data is quite consistent with the random
model containing no intrinsic frequencies apart from the 11-year
cycle.  
 Note that the probability of 0.253 for a $3\sigma$ peak
  is independent of the null case used for determining $\sigma$:
  it is based on a $\sim$0.00135 probability of finding a $3\sigma$ peak
  at any given resolved frequency bin, and having 216 resolved frequency
  bins.

\begin{figure}
	\includegraphics[width=\columnwidth]{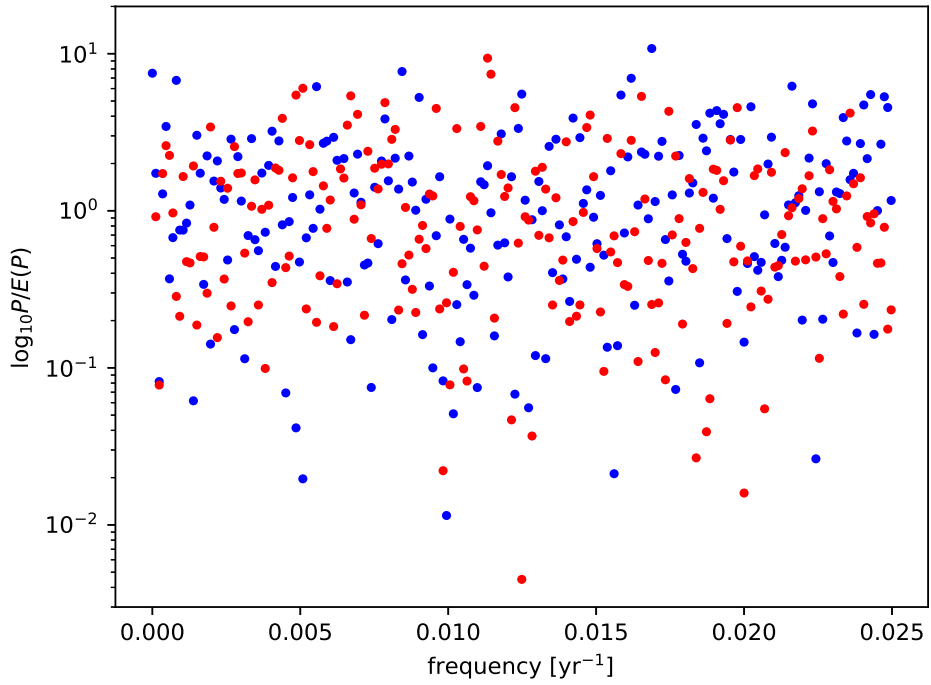}
        \caption{Ratio of the power in a frequency bin divided by the
          median power of the NNF model realizations at that
          frequency. The blue points correspond one realization of the
          noisy normal form model, the red points to the reconstruction
          from cosmogenic isotopes (cf. Fig.~\ref{fig:PS}).}
    \label{fig:errs}
\end{figure}

\begin{figure}
	\includegraphics[width=\columnwidth]{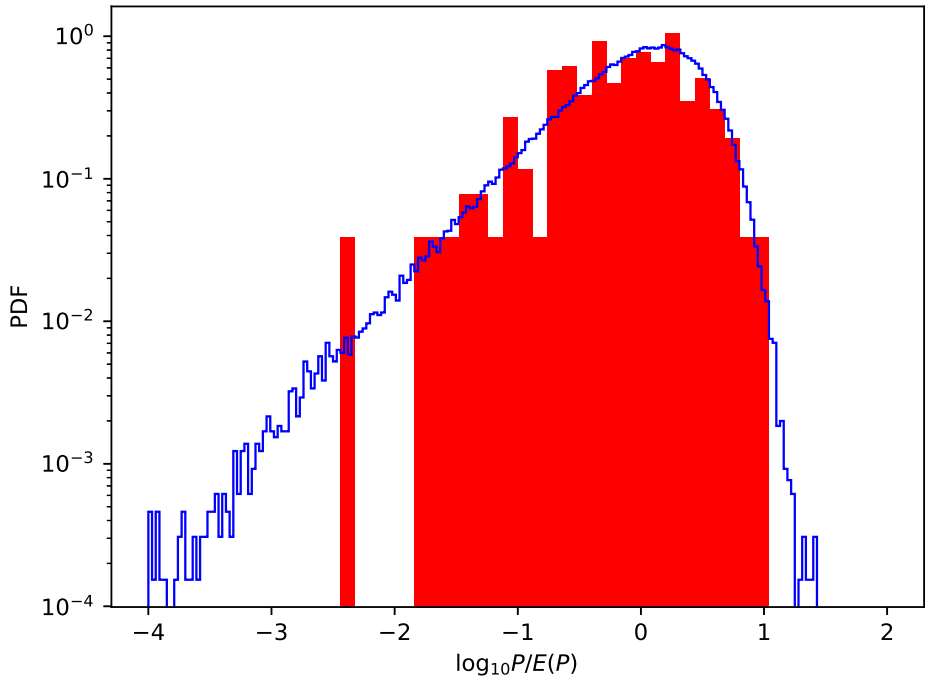}
        \caption{Probability distribution function of the ratio of the
          power in a frequency divided by the median power of the NNF
          model realizations at that frequency. The blue curve is the
          expectation value based on 1000 realizations of the model, the
          red bars correspond to the reconstruction from cosmogenic
          isotopes.}
    \label{fig:pdf}
\end{figure}

As a further comparison of the empirical and model power spectra, we
consider the scatter with respect to the expectation value (median of
the 1000 NNF model realizations) shown in Fig.~\ref{fig:errs}.  
Application of the Kolmogorov-Smirnov test shows that the
  distributions for the empirical data and for the NNF model (red and
  blue dots, respectively) are consistent with each other at the
  $2\sigma$ (95\%) level.
This is further illustrated by Figure~\ref{fig:pdf}, which shows
probability distribution functions (PDF) of the ratio between the
power at a given frequency and the median of the model realizations,
considering all periodicities above 40 years. 
 The consistency of the scatter in the empirical power spectrum
  with the NNF model demonstrates that the model captures the noise
  statistics of the empirical data quite well.

\begin{figure}
	\includegraphics[width=\columnwidth]{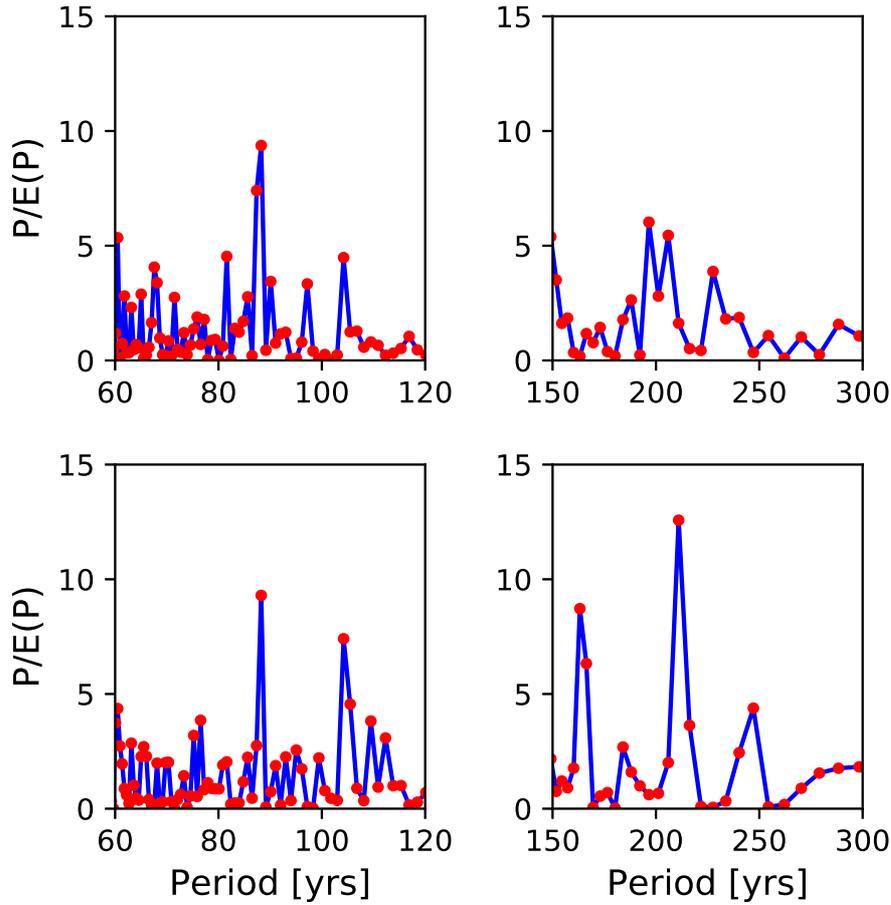}
        \caption{Segments from power spectra around the Gleissberg (left
          panels) and the de Vries (right panels) periodicities. Shown
          is the ratio of power to the expectation value (median of 1000
          realizations of the NNF model).  Spectra from reconstruction
          based upon cosmic isotopes by (upper panels) are compared with
          those from selected realizations of the NNF model with
          $3\sigma$ peaks in the same range of periods (lower
          panels). The red dots emphasize the frequency resolution. The
          peaks from the empirical data are similarly sharp as those
          from the model, indicating the random origin of the former.}
    \label{fig:peaks}
\end{figure}

Another property that sheds light upon the nature of peaks in the power
spectra is their sharpness.  The $3\sigma$ peaks appearing in the NNF
model are usually quite narrow (at most a few resolution elements in
frequency). This results from the fact that the random noise in the
model is uncorrelated in frequency space. On the other hand, an
intrinsic periodicity of the dynamo process is expected to yield a
broader peak (such as the 11-year band) since the noise effectively
leads to a finite coherence time of the dynamo
modes. Fig.~\ref{fig:peaks} demonstrates that the peaks corresponding to
the observed Gleissberg and de Vries periodicities indeed are quite
sharp, similar to those resulting from the NNF model.

\subsection{Analysing subseries}

It has been argued that the appearance of certain periods during
different stretches of time throughout the interval covered by the
cosmogenic isotope data would provide evidence against a random origin
\citep[e.g.,][]{2018arXiv181011952S}. This can easily be tested using
the NNF model as a null case. 

\begin{figure}
  \includegraphics[width=\columnwidth]{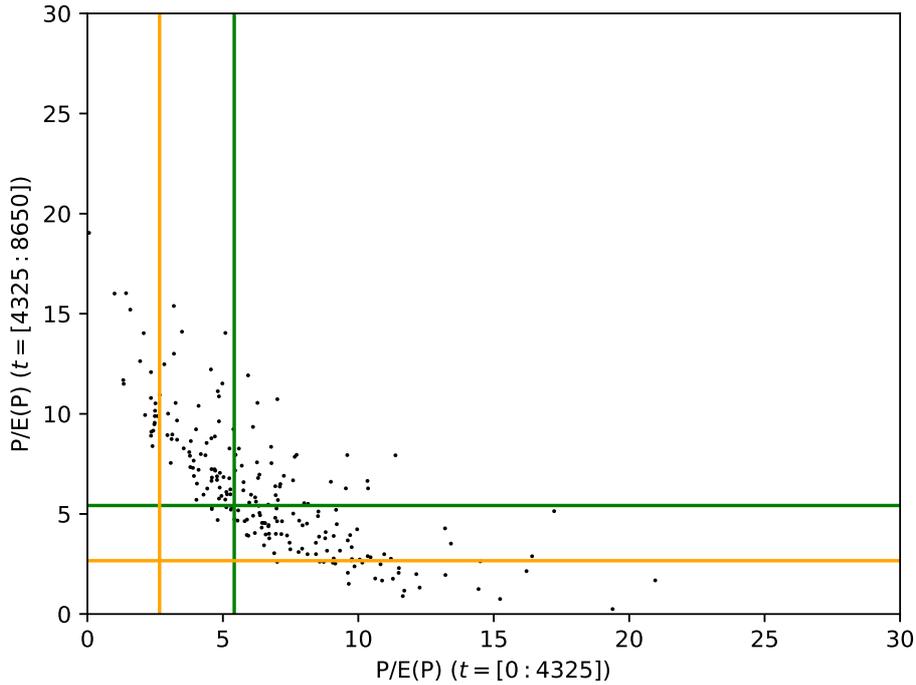}
  \caption{Period analysis of subseries from realizations of the NNF
    model. Each point in the diagram corresponds to one of the 216
    highest peaks with periods above 40 years in the normalized power
    spectra (power divided by the expectation value given by the
    median) from 1000 realizations. They are plotted according to the
    normalized power at the same period in the first half (horizontal
    axis) and in the second half (vertical axis) of the 10,000 years
    full series. The distribution of the points show that significant
    power in the full series in most cases corresponds to high power
    in both subseries.  84\% of all points in the power spectrum from
    the two subseries lie beneath (to the left of) the orange
    horizontal (vertical) line and 97.7\% lie beneath (to the left of)
    the green lines.}
  \label{fig:P1P2}
\end{figure}

For 1000 realizations of the NNF model covering 8650 years each,
  we applied the Lomb-Scargle periodogram method and selected the 216
  realizations with the highest (3-sigma) peaks. For the thus selected
  realizations and peaks, we determined the spectral power at the
  frequencies corresponding to these peaks when the first and second
  halves of the corresponding time series were analyzed separately.
  The values of the normalized power thus obtained for the first and
  second halves of the random series are plotted against each
  other in Fig.~\ref{fig:P1P2}. The resulting distribution shows that
  in most cases of a significant spectral peak in the full series,
  there is significant power as well for both half series. Since all
  peaks in the NNF model (without intrinsic periodicities apart from
  the 11-year cycle) are due to stochastic noise, this demonstrates
  that the consistent appearance of high peaks in both subseries does
  not per se provide evidence for a non-random (intrinsic) nature of
  such peaks.  It is clear that there is always a better chance for a
  significant peak in the full series if it is already strong in both
  half series.

\subsection{Bandpass filtering}

Bandpass filtering of the power spectrum in the range of the Gleissberg
and de Vries periods and back transformation into the time domain has
been used, e.g., by \cite{Beer:etal:2018} in an attempt to elucidate the
nature of the periodicities found in the empirical data.  However,
applying the same procedure to realizations of the NNF model shows that
bandpass filtering does not provide any evidence for the intrinsic
nature of these periodicities.

\begin{figure}
  \includegraphics[width=\columnwidth]{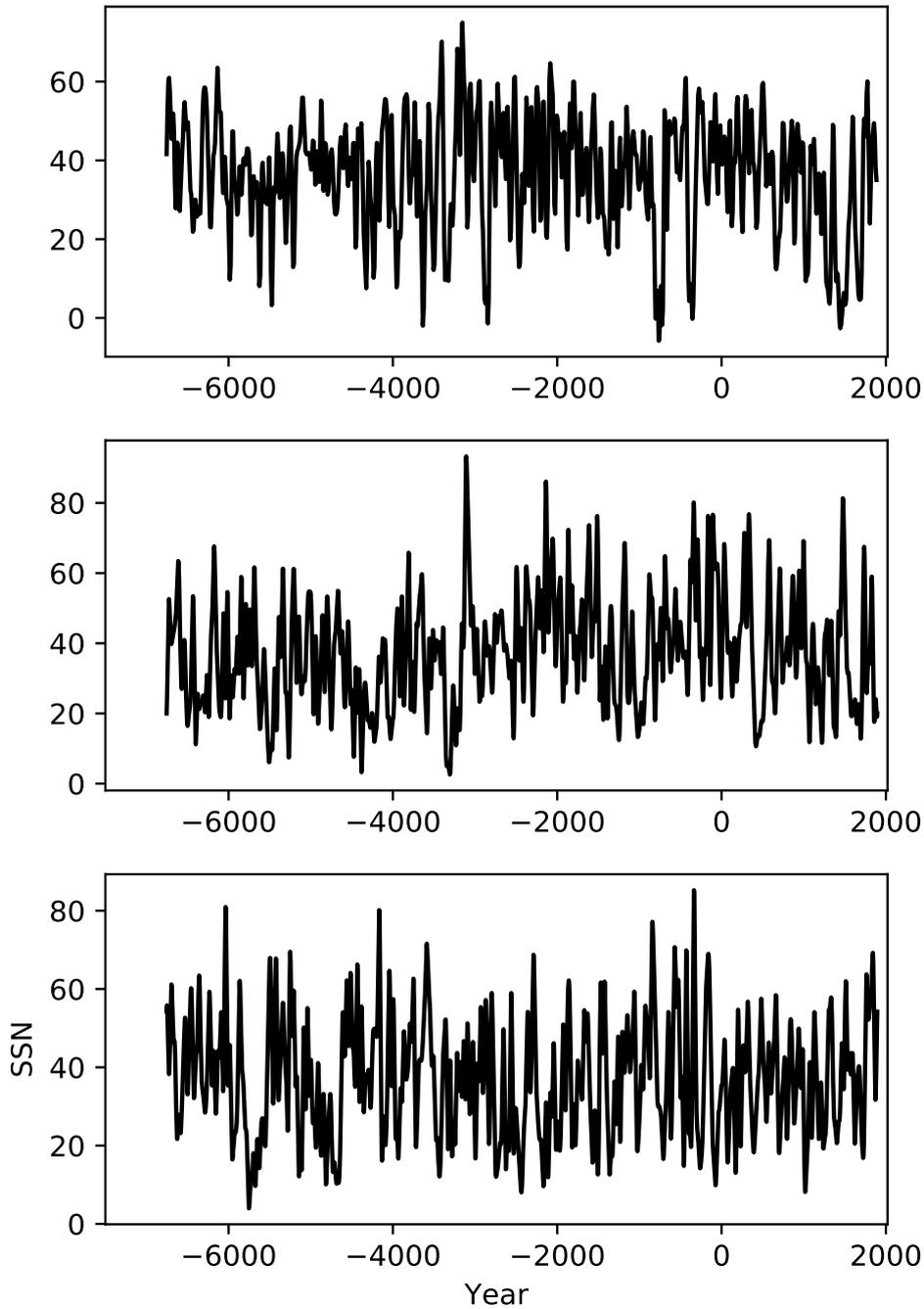}
  \caption{Long-term activity records. {\it Top panel:}
    reconstructed sunspot number based on cosmogenic isotopes
    \citep{Usoskin:etal:2016}. {\it Middle panel:} realization of the
    noisy normal form model with a $3\sigma$ peak in the range of the
    Geissberg period. {\it Bottom panel:} realization of the
    noisy normal form model with a $3\sigma$ peak in the range of the
    de Vries period (cf. Fig.~\ref{fig:peaks}).}
  \label{fig:realizations}
\end{figure}

We consider the two realizations of the NNF model that show peaks in
the range of the Gleissberg and de Vries periodicities
(cf. Fig.~\ref{fig:peaks}). The time evolution of the sunspot number
(SSN) for these realizations is shown in Fig.~\ref{fig:realizations},
together with the corresponding record of the reconstructed sunspot
number based on cosmogenic isotopes \citep{Usoskin:etal:2016}.
Periods below 40 years were filtered out in all three time series.

\begin{figure}
  \includegraphics[width=\columnwidth]{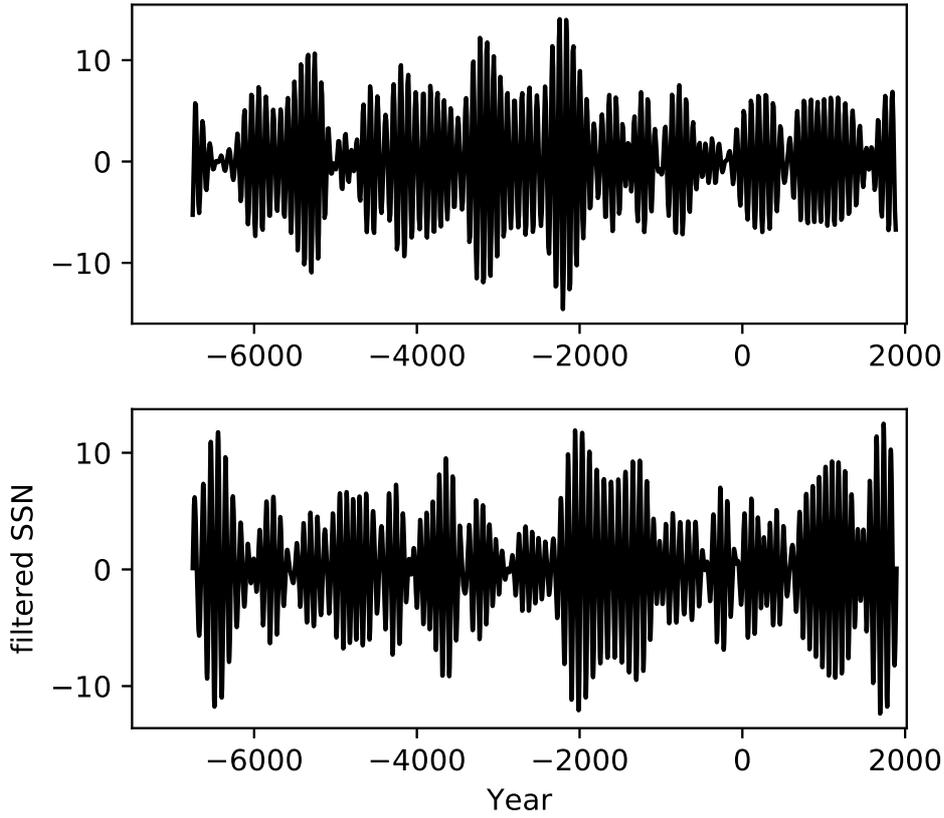}
  \caption{Passband filtered time evolution of the sunspot number in the
    period range between 75~yr and 100~yr around the Gleissberg period.
    The result from the empirical record reconstructed from cosmogenic
    isotopes (upper panel) are shown in comparison to that from a
    realization of the noisy normal form model (lower panel).}
  \label{fig:Gleissberg}
\end{figure}

\begin{figure}
  \includegraphics[width=\columnwidth]{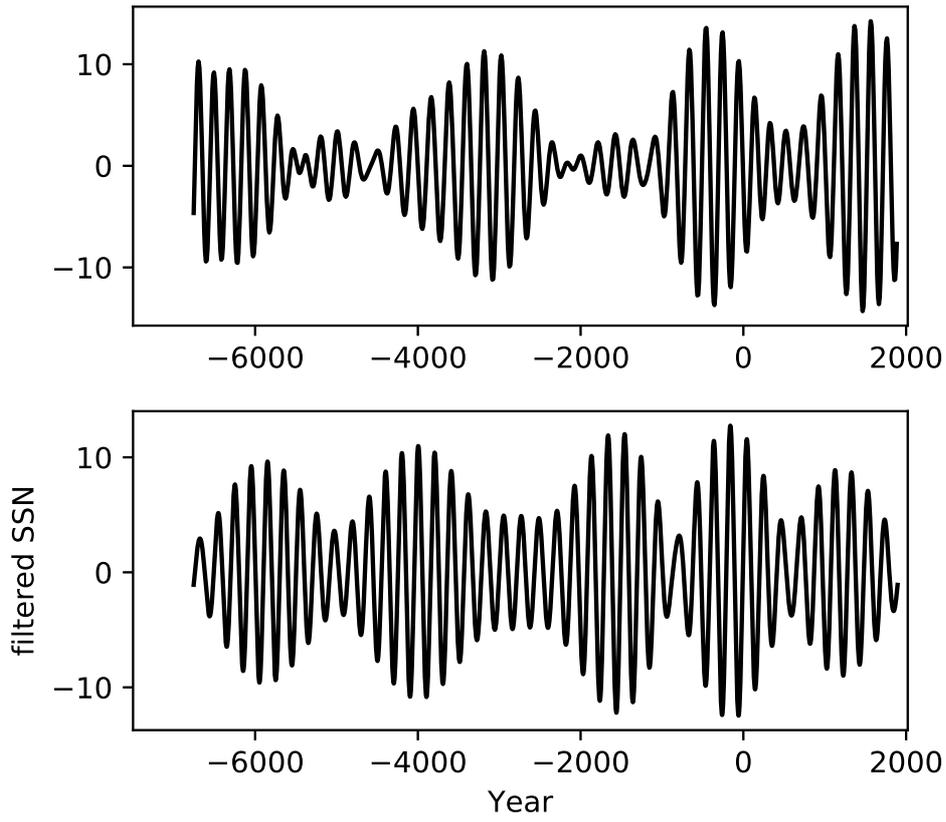}
  \caption{Same as Fig.~\ref{fig:Gleissberg} with passband filtering in
    the range between 180~yr and 230~yr around the de Vries period .}
  \label{fig:deVries}
\end{figure}

Figures~\ref{fig:Gleissberg} and \ref{fig:deVries} show, respectively,
the results of passband filtering in the ranges 75--100~yrs (Gleissberg)
and 180--230~yrs (de Vries) in comparison with the
corresponding pattern obtained by applying the same filtering to the
empirical record.  The patterns from the NNF model and from the
real data are qualitatively similar, suggesting that the corresponding
frequency (Gleissberg and de Vries) seen in the cosmogenic isotope
record is consistent with a random process.

\begin{figure}
  \includegraphics[width=\columnwidth]{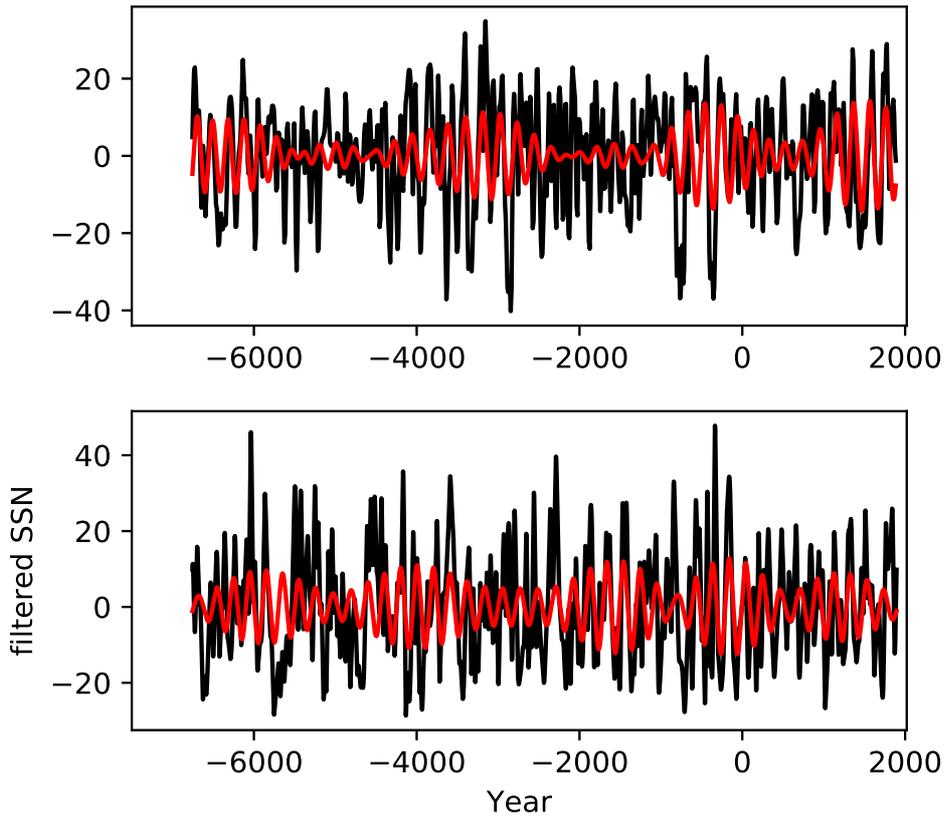}
  \caption{High-pass filtered records of the empirical (upper panel)
    and modelled sunspot numbers (lower panel). Overplotted in red are
    the corresponding bandpass-filtered records in the range of the de
    Vries period (cf. Fig.~\ref{fig:deVries}.) }
  \label{fig:Beer-fig5}
\end{figure}

Finally, Fig.~\ref{fig:Beer-fig5} shows the empirical record after
bandpass filtering in the range 40--1000 years (black curve) and
180--230 years (around the de Vries perio, red curve), together with
the result after applying the same procedures to the realization of
the NNF model. This figure can be compared to the analogous Fig.~5 of
\citet{Beer:etal:2018}. The extrema of the filtered de Vries
periodicities tend to align with some of the extrema (grand minima and
grand maxima) of the original series, giving the impression of a
periodic appearance of such extrema. However, this is also the case in
the NNF series with a random excitation of a periodicity in the same
range.

Since bandpass filtering of the NNF realizations leads to qualitatively
similar results as applying the same procedure to the empirical data, we
conclude that such filtering does not provide evidence for the intrinsic
nature of the related periodicities.

\section{Conclusion}

Our analysis shows that the fluctuations in the power spectrum of the
sunspot numbers reconstructed from cosmogenic isotopes are consistent
with a weakly nonlinear and noisy limit cycle with no intrinsic
periodicities except that of the basic 11/22-year cycle.  Such a mode of
operation of the solar dynamo is suggested by solar and stellar
observations and can be faithfully described by a generic noisy normal
form model with parameters taken from observations
\citep{Cameron:Schuessler:2017}. Seemingly significant periodicities
such as the $\sim 90$-year Gleissberg and the $\sim 210$-year de Vries
``cycles'' are expected to occur in random realizations as a result of
the stochastic noise in the dynamo excitation. This conclusion is
further strengthened by the sharpness of the corresponding peaks in the
power spectrum, indicating a random origin.

Of course, our analysis cannot per se exclude that these periodicities
may be intrinsic after all, but we have shown that, so far, this notion
is not supported by the data, which are consistent with the NNF null
case. The interpretation of such periodicities in terms of dynamo theory
\citep[e.g.,][]{Beer:etal:2018} should therefore be considered with due
caution.

\section*{Acknowledgements}

I. G. Usoskin kindly provided the sunspot number reconstruction from the
cosmogenic isotope record presented in \cite{Usoskin:etal:2016}.

\bibliographystyle{aa}
\bibliography{dynamo} 

\newpage
\newpage

\newpage

\newpage

\newpage

\newpage

\newpage

\newpage

\newpage

\label{lastpage}
\end{document}